\documentclass[preprint]{elsarticle} \usepackage{graphicx} \usepackage{epsfig}
\usepackage{amsmath} \usepackage{amssymb} \usepackage{amsfonts} \usepackage{color}
\begin{document}
\begin{frontmatter} \title{Entropy and equilibrium state of free market models} \author[ppge,if]{J. R. Iglesias} \ead{roberto@if.ufrgs.br} \author[if]{R.
M. C. de Almeida} \ead{rita@if.ufrgs.br} \address[ppge]{Programa de P\'os-Gradua\c{c}\~ao
em Economia Aplicada, UFRGS, Av. Jo\~ao Pessoa 52, 90040-000 Porto Alegre, RS, Brazil}
\date{\today} \address[if]{Instituto de F\'{\i}sica, UFRGS, and Instituto Nacional de
Ci\^encia e Tecnologia de Sistemas Complexos, Caixa Postal 15051, 91501-970 Porto Alegre,
RS, Brazil.}

\begin{abstract} Many recent models of trade dynamics use the simple idea of wealth exchanges among economic agents in order to obtain a stable or equilibrium distribution of wealth among the agents. In particular, a plain analogy compares the wealth in a society with the energy in a physical system, and the trade between agents to the energy exchange between molecules during collisions. In physical systems, the energy exchange among molecules leads to a state of {\it equipartition} of the energy and to an equilibrium situation where the entropy is a maximum. On the other hand, in the majority of exchange models, the system converges to a very unequal {\it condensed} state, where one or a few agents concentrate all the wealth of the society while the wide majority of agents shares zero or almost zero fraction of the wealth. So, in those economic systems a minimum entropy state is attained. We propose here an analytical model where we investigate the effects of a particular class of economic exchanges that minimize the entropy. By solving the model we discuss the conditions that can drive the system to a state of minimum entropy, as well as the mechanisms to recover a kind of equipartition of wealth. \end{abstract}


\end{frontmatter}

\section{Introduction}
The second law of thermodynamics states that isolated systems always tend to an equilibrium state of maximum entropy, where equilibrium means that the macroscopic properties of the system are the same in any part of it. The second law of thermodynamics can also be deduced from an analysis of the efficiency of a thermal engine, and according to Clausius: ``No process is possible whose sole result is the transfer of heat from a body of lower temperature to a body of higher temperature''~\cite{Reichl} In modern statistical physics the second law may be inferred from Boltzmann H-theorem~\cite{Reichl}
and it is generally accepted as a natural consequence of the energy exchange among
molecules in the kinetic theory of gases: when molecules collide in a gas there is a transfer of kinetic energy from the more energetic to the less energetic molecules and, as a consequence, after a transient, all molecules share, in average, the same energy. This result is
the theorem of equipartition of energy that characterizes the Maxwell-Boltzmann
equilibrium distribution of velocities, and it is associated with the maximum entropy
state. However, in particular situations, it may happen that the entropy decreases as a
function of time; examples are self-organized systems and biological organisms. One common
characteristic of those systems is that they are not in equilibrium even if they seem stable or stationary in time~\cite{Bak}.

The idea that economic systems also possess some kind of equilibrium state is an underlying
concept in classical economic theory, going from Pareto optimality to Nash  equilibrium~\cite{masco}. The Edgeworth box model is an example of how a simple exchange conservative model can lead two consumers to optimize their respective utility~\cite{masco}. But in other cases there are symptomatic evidences that economic systems can be out of equilibrium and/or exhibit metastable equilibrium states. This is the case when studying the wealth and income distribution. Fluctuations around an equilibrium state behave in a Gaussian or normal way: The probability of rare events is very small, as the Gaussian distribution exhibits exponential tails. On the other hand, empirical studies focusing the income distribution of workers, companies and countries were first presented more than a century
ago by Vilfredo Pareto and he discovered that the income distribution does not behave
in a Gaussian way but exhibits ``heavy tails'', i.e. the cumulative probability $P(w)$ of
workers whose income is at least $w$ follows a power law~\cite{Pareto} given by $P(w)
\propto w^{-\beta}$. Nowadays, this power law distribution is known as Pareto
distribution, and the corresponding exponent $\beta$ is named Pareto exponent. However, recent data
indicates that, even though Pareto distribution provides a good fit in the high
income range, it does not agree with the observed data over the middle and low income
range. For instance, data from Japan~\cite{souma,nirei}, Italy~\cite{clementi},
India~\cite{sinha1}, the United States of America and the United
Kingdom~\cite{dragu2000,dragu2001a,dragu2001b} are fitted by a log-normal or Boltzmann
distribution with the maximum located at the middle-income region plus a power law for the high-income strata. The existence of these two regimes may be justified in a qualitative
way by stating that in the low and middle income classes the process of wealth accumulation
is additive (and mainly due to wages), causing a Gaussian-like distribution, while in the high income range, wealth grows in a multiplicative way, generating the observed power law
tail~\cite{nirei}.

In recent years physicists and economists working in complexity science proposed different
mathematical models of wealth exchange among economic agents in order to try to explain
these empirical data (For a review see refs.~\cite{Yakov,chatter3,Caon2007}). Quoting ref. \cite{Yakov}: {\it Inspired by Boltzmann's kinetic theory of collisions in gases, econophysicists introduced an alternative two-body approach, where agents perform pairwise economic transactions and transfer money from one agent to another. Actually, this approach was pioneered by the sociologist John Angle~\cite{Angle86,Angle93}}. Most of these
models consider an ensemble of interacting economic agents, each one possessing a given
amount of endowments, money\cite{dragu2000} or assets\cite{Caon2007,chakra} that represents its
economics resources and that we will describe as ``wealth''. Most of these models focus on one particular aspect of economic processes: The competition among different agents (countries, enterprises, etc.) acting in an environment where all exchanges of wealth between agents take place in a conservative manner, i.e., a conservative exchange market model (CEMM)˜\cite{PIAV2003}.
This restriction has several motivations: On the one hand, it can be argued that resources
are material objects, and consequently they cannot be created or destroyed by means of
exchanging them. On the other hand, the use of the CEMM implies that the
exchange model is a zero-sum game, something that may seem at odds with usual economic
orthodoxy. However, the results also hold for systems in which the total amount of wealth
increases uniformly and smoothly in time. The interaction among agents consists in a exchange of a fixed~\cite{dragu2000} or random~\cite{Caon2007,chakra} amount of their wealth. The process of exchange is similar to the collision of molecules in a gas and the
amount of exchanged wealth when two agents interact corresponds to some economic ``energy''
that may be transferred for one agent to another. If this exchanged amount corresponds to a
fixed or random fraction of one of the interacting agents wealth, the resulting wealth
distribution is -- unsurprisingly -- a Gibbs exponential distribution~\cite{dragu2000}. 

Aiming at obtaining distributions with power law tails, in order to describe the higher
income region of the wealth distribution histogram, several methods have been proposed
mostly introducing a multiplicative risk aversion that acts as a multiplicative noise.
Numerical results\cite{Caon2007,chakra,sinha2,chatter1,chatter2,IGPVA2003,IGVA2004}, as
well as some analytical calculations~\cite{Bouchaud,cristian07}, indicate that a frequent
outcome in these models is {\it condensation}, i.e. concentration of all available wealth
in just one or a few agents. This final state corresponds to a kind of equipartition of poverty: 
All agents (except for a set of zero measure) possess zero wealth while one, or a few ones, concentrate all available resources. In any case the final configuration is a stationary state of ``equilibrium'', since agents with zero wealth cannot participate in further exchanges. Several methods have been
proposed to avoid this situation, for instance, exchange rules where the poorer agents are
favored~\cite{sinha1,Caon2007,IGVA2004,cristian07,west} or taxes and
regulations~\cite{ausloos,igles2010}. Here, instead, we are mainly interested in the
condensed state and in the dynamics driving the system to this condensed state, as well as
in the entropy behavior when the system approaches condensation.

As previously stated, exchange rules are determinant of the long-time behavior of the system. While we can obtain an exponential Boltzmann-Gibbs distribution - and consequently a maximum entropy state - if the exchanged fraction is fixed or determined at random, condensation is the outcome when exchange rules are so that, when two agents interact, the exchanged amount $\Delta w$ is proportional to the wealth of one of the participants or to
both~\cite{Bouchaud} but in no case one participant can win more that the value he put in
stake. So, the exchange process is a kind of lottery where no agent can win more than his own possessions. One particular and widely used exchange rule is to consider that the fraction of transferred wealth from agent $x'$ to agent $x''$, or vice versa, is: $\Delta w = \mbox{min} \{ (1-\lambda)w';(1-\lambda)w''\}$,
where $w'$ and $w''$ are the respective wealth of the two interacting agents, and $\lambda$ is a risk-aversion factor, so the capital fraction that the agents risk during the 
exchange is $1-\lambda$~\cite{Caon2007,IGVA2004,cristian07}. It is worth noting that even approaching a condensed state, in the intermediate stages the wealth distribution goes through a series
of power law distributions where the Pareto exponent increases as a function of
time~\cite{cristian07}. The problem with the previous definition of $\Delta w$ is that it involves a logical
comparison that is difficult to be treated analytically. Here we consider another form of
$\Delta w$. We define an analytical expression of $\Delta w$ that guarantees that no agent
participating in the exchange risk more than the quantity he can win. To do that we define
$\Delta w = \frac{w' w''}{w' + w''}$, that presents similar properties and is equal to the
wealth of the poorer partner when the wealth of the richer agents is much bigger than the
other. Notice also that we eliminated the risk-aversion factor, just by considering $\lambda=0$, because we have verified that one does not need the multiplicative noise to obtain the condensed state. Using this exchange rule and the methods of non-equilibrium statistical mechanics we show in this paper that the entropy decreases in the intermediate stages leading to a condensed state of minimum entropy (and maximum inequality, Gini coefficient equal to 1).

The text is organized as follows: in the next section we write the exchange model in the
form of a master equation that is solved by numerical iteration in section 3. Next in
section 4 we calculate the entropy of the system as well as the Theil coefficient and we
discuss why in this case the Boltzmann H-theorem is not verified. Finally the results are
discussed and the conclusions presented in section 5.

\section{The evolution equation}
We consider a collection of $R$ individuals, each one characterized by a given value of a continuous
variable $w$. This variable can represent the wealth, cash, properties, or some other
measure of the agent's fortune, but it can also represent a physical scalar quantity, as
energy. Like in the kinetic theory of gases, agents may interact and exchange any fraction
of $w$. In the standard kinetic theory the exchange of random fractions of energy leads to
an equilibrium state described by Maxwell-Boltzmann distribution\cite{Reichl}. In the same
way, considering $w$ as representing the money owned by an agent, and random exchanges
between agents, a maxwellian money distribution is also
found\cite{dragu2000,dragu2001a,PIAV2003,SI2004}. However, as previously stated, different
exchange mechanisms can be considered. Several of them are in some way proportional to the
wealth owned by the participating agents and leads to a condensed
state\cite{Caon2007,SI2004,chatter1,chatter2,IGPVA2003,IGVA2004,Bouchaud}. Here we study a
very simple non-linear exchange rule. Let us consider two agents, with wealth $w'$ and
$w''$, respectively. We assume that when they interact they exchange $\Delta w=(w'w'')/(w'
+ w'')$. It implies that when the two agents have very similar wealth, the exchanged amount
is approximately half of each agent wealth, while if the possessions are very different, the
transferred amount is equal to the poorer agent's wealth. This expression yields a fair
exchange rule in the sense that no agent risks more that the amount he can win. Indeed the
rule is very similar to the one used in refs.~\cite{Caon2007,IGVA2004}, i.e.
$w=\mbox{min}\{w,w'\}$ but here we avoid the logical operator using an analytical
expression.

We also aim at investigating the effect of favoring the poorer agent in an interaction.
Condensation of wealth is an undesirable effect in models describing wealth distribution
and many authors proposed different ingredients to avoid this effect. One of the most popular recipes is to define a wealth-dependent probability of winning a transaction. For
example in refs.~\cite{sinha1,Caon2007,IGVA2004,cristian07,west} it is assumed that the
winning probability is higher for the poorer participating agent. Even if this recipe seems
counterintuitive, favoring the poorer agent in every transaction somehow emulates a
regulatory policy imposed by a government. To assess the effects of this kind of policy,
we assume that in an interaction between two agents with wealth $w'$ and $w''$, the
probability $p_{\gamma}(w'|w'')$ that the agent with wealth $w'$ wins the exchanged amount,
$\Delta w=(w' w'')/(w' + w'')$, is given by \begin{equation} \label{eq:gain}
p_{\gamma}(w'|w'')= \frac{1}{2} \left[1 - \gamma \tanh\left(w'-w''\right)\right],
\end{equation} where $\gamma$ may assume any value in the interval $[0,1]$. This winning
probability has the following properties \begin{itemize} \item
$p_{\gamma}(w'|w'')=\frac{1}{2}$, if $w'=w''$ for all $\gamma$. \item
$p_{\gamma}(w'|w'')=\frac{1}{2}$, if $\gamma=0$ for all $w'$ and $w''$. \item
$p_{\gamma}(w'|w'')=\frac{1}{2}(1 \mp \gamma)$, if $w'-w'' \rightarrow\pm\infty$, so $0
\leq \gamma \leq 1$. \item $p_{\gamma}(w'|w'')+p_{\gamma}(w''|w')=1$, for all $0 \leq
\gamma \leq 1$, $w'$, and $w''$. \end{itemize} Hence, by varying $\gamma$ we can
investigate the effects of a varying the winning probability on the wealth distribution.

We define $N(w,t) dw$ as the probability of finding an agent with wealth in the interval
$[w,w+dw]$ at time $t$ or, alternatively, as the relative number of agents with wealth in
that interval. From now on, we consider the limit where there is an infinite number of agents, such that $N(w,t)$ is a continuous function of $w$. The wealth evolution of such
agents depend on their transactions. Assuming that
{\it i)} the probability per unit time that two agents of wealth $w$ and $w'$ perform a
transaction is given by $k \; N(w,t)\,N(w',t)\,dw\, dw'$, with $k$ being the transaction
frequency;
{\it ii)} the exchanged wealth in that transaction is $\Delta w=(w'w'')/(w' + w'')$; and
{\it iii)} the probability of gaining or loosing that amount is given by a regulatory
function $p_{\gamma}(w'|w'')$ given by Eq.(\ref{eq:gain}); we may write the evolution
equation for $N(w,t)$ as \begin{eqnarray} \label{evol1} && N(w,t+\Delta t) - N(w,t) =
-2\;k\;N(w,t)\nonumber \\ && + k\; \int_0^{\infty}{dw'}
\int_0^{\infty}{dw''}N(w',t)N(w'',t) \left\{p_{\gamma}(w'|w'')\; \delta \left( w-\left[
w'+\frac{w'\;w''}{w'+w''}\right] \right)\right. \nonumber \\ && + \left.
p_{\gamma}(w''|w')\; \delta\left( w-\left[ w'-\frac{w'\;w''}{w'+w''}\right] \right)+
p_{\gamma}(w''|w')\;\delta\left( w-\left[ w''+\frac{w'\;w''}{w'+w''}\right] \right)\right.
\nonumber \\ && + \left. p_{\gamma}(w'|w'')\; \delta\left( w-\left[
w''-\frac{w'\;w''}{w'+w''}\right] \right) \right\} . \end{eqnarray} \noindent This is a
non-linear probability conservation equation. The first term in the right hand side of
Eq.(\ref{evol1}) gives the amount of agents that have changed from $w$ to some other value.
This term is proportional to the total number $K(w)$ of transactions involving agents with
wealth in the range $(w,w+dw)$, during the time interval $dt$, which is given by
\begin{eqnarray} \label{eq:K1} K(w)&=&k \int_0^{\infty}{dw'}
\int_0^{\infty}{dw''}N(w',t)N(w'',t)\left[\delta(w'-w)+\delta(w''-w)\right]\\ &=&
2\;k\;N(w,t)\nonumber , \end{eqnarray} and where we have explicitly taken into account the
normalization of $N(w,t)$, that is, $\int_0^{\infty}{dw}N(w,t)=1$. The other four terms in
Eq.(\ref{evol1}) describe the increment in $N(w,t)$ due to agents that changed their wealth
to $w$ (by winning money, first and third terms, or by losing money, second and fourth
terms) during $dt$. The conservation of number of agents is easily verified by integrating
Eq.(\ref{evol1}) in $w$. The total wealth is also conserved. As the number of agents is
conserved, wealth conservation is demonstrated by showing that the time derivative of the
average wealth is zero: \begin{equation} \frac{\mbox{d}}{\mbox{d}t} \langle w
\rangle=\int_{0}^{\infty} dw\; w\frac{\partial N(w,t)}{\partial t}= 0, \end{equation} Using
Eq.(\ref{evol1}) to estimate the time derivative of $N(w,t)$ and performing the integrals
containing $\delta$-functions we obtain \begin{eqnarray} && \int_0^{\infty} dw \; w \left[
N(w,t+\Delta t)-N(w,t)\right]= \nonumber \\ && -2k\int_0^{\infty} dw \; w
N(w,t)+\int_0^{\infty}{dw'} \int_0^{\infty}{dw''}N(w',t)N(w'',t) k \times \nonumber \\ &&
\left\{p_{\gamma}(w'|w'') \left[ w'+\frac{w'\;w''}{w'+w''}\right]+p_{\gamma}(w''|w')\left[
w'-\frac{w'\;w''}{w'+w''}\right]\right. \nonumber \\ && \left. +p_{\gamma}(w''|w')\left[
w''+\frac{w'\;w''}{w'+w''}\right] + p_{\gamma}(w'|w'')\left[
w''-\frac{w'\;w''}{w'+w''}\right]\right\} = 0 , \end{eqnarray} implying a constant average
wealth. We use wealth and number of agents conservation to define the wealth unit as the
average wealth given by $\langle w \rangle =
\int_0^{\infty}{dw}w\;N(w,t)/\int_0^{\infty}{dw}N(w,t)$. In what follows wealth is given in
units of average wealth $\langle w \rangle$.

\section{Numerical iteration}
\begin{figure} \begin{center}
\includegraphics[width=7cm]{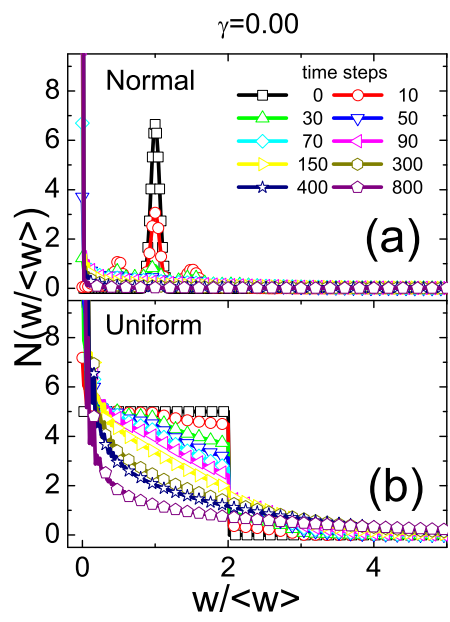} \end{center} \caption{Plot of the wealth
distribution at different times for (a) an initial normal distribution, (b) an initial
uniform distribution.} \label{fig:Nwlin} \end{figure}

\begin{figure} \begin{center} \includegraphics[width=8cm]{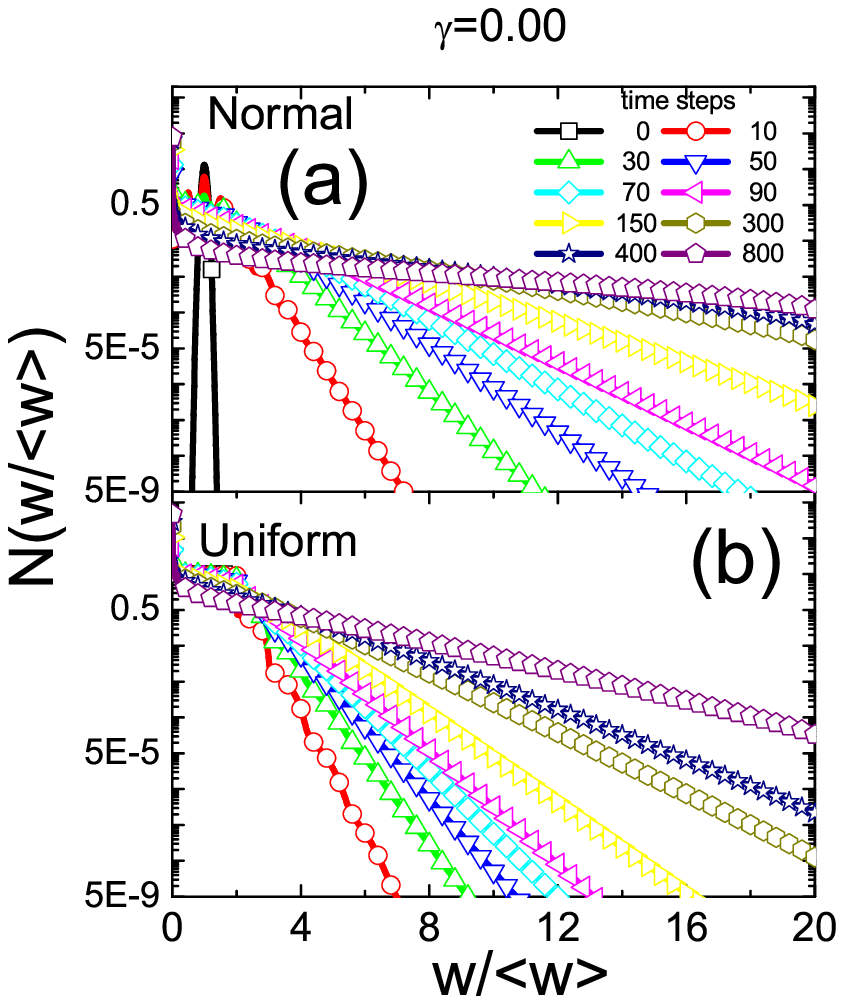} \end{center}
\caption{Log-linear plot of the wealth distribution at different times for (a) an initial
normal distribution, (b) an initial uniform distribution.} \label{fig:Nwlog} \end{figure}

The evolution equation, Eq.\ref{evol1}, has been numerically iterated by dividing the $w$
axis in bins of width $<w>/100$, while the exchange rate $k$ is taken as $k=0.01/\tau$ with
$\tau$ being the simulation time step. We have considered two different initial conditions:
\begin{itemize} \item a normal distribution with average $\langle w \rangle$ and variance
$\langle w \rangle/6$. \item an uniform distribution in the interval $[0,2\langle w
\rangle]$. \end{itemize}
We first analyze the symmetric case, where $\gamma=0$, such that in each transaction the
winning probability is 1/2 for both agents. Fig.~\ref{fig:Nwlin} shows the time evolution
in both cases, Fig.~\ref{fig:Nwlin}(a) for a normal initial distribution and
Fig.~\ref{fig:Nwlin}(b) for a uniform initial distribution. It is interesting to note that
after some initial oscillations, more pronounced for the case of the normal distribution
(because at the beginning of the simulation the exchanged amounts are always very near
$<w>/2$) the system arrives to the same distribution. This final distribution evolves to
the condensed state. It is characterized by a very small fraction of agents with wealth
above $<w>$, and an ever increasing number of agents with wealth very near zero.
However, it is worth to note that the transient states present a clear exponential tail
(and not power law), as one can see in the log-linear plots presented in
Fig.~(\ref{fig:Nwlog}, where the distributions are represented by a linearly decreasing
function with an slope, $-\zeta$, that corresponds to the exponent of the distribution.
This exponent decreases with time as shown in Fig. (2). One expects that for $t \rightarrow
\infty$ the exponent goes to zero, indicating that all the agents have zero wealth.
Nevertheless, as the total wealth is conserved a finite set of agents (one or a few, but in
any case a zero-measure set) possess all the available resources On the other hand, the
interval on the $w$ axis where $N(w,t)$ is described by an exponential increases with time,
and ranges from $w$ of the order of $\langle w \rangle$ to a higher value of $w$ that
increases in time. After that, $N(w,t)dw $ decreases very fast to values much smaller than
$10^{-9}$ implying that it is very unlikely that a system described by that $N(w,t)$ would
present agents in the high wealth interval. We call this distribution, where the number of
agents goes to infinity for $w$ going to zero, and goes to zero for any finite wealth, an
$L-$shaped distribution. 

\begin{figure} \begin{center} \includegraphics[width=8cm]{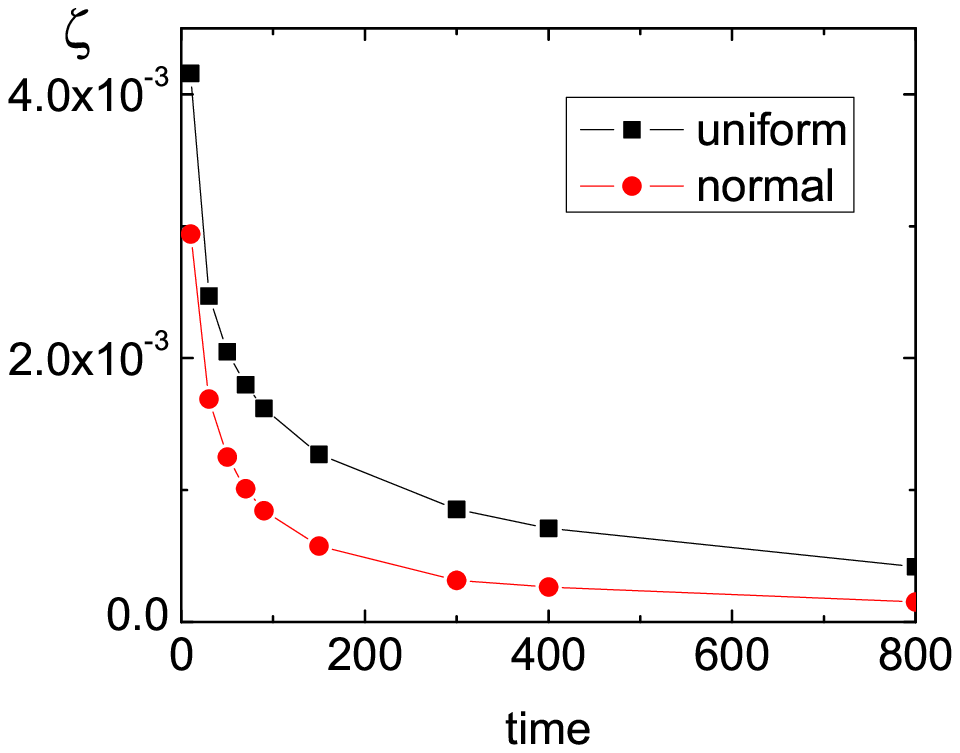}
\label{fig:exponindex0} \end{center} \caption{The exponent of the wealth distribution as a
function of time for different initial conditions. } \end{figure}

\begin{figure} \begin{center} \includegraphics[width=8cm]{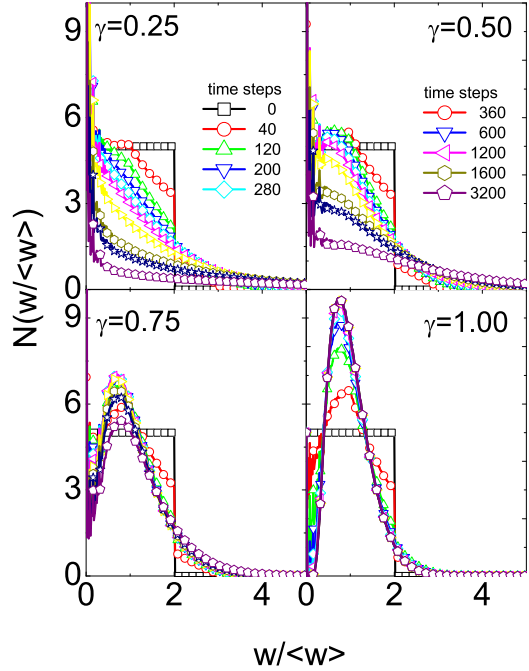} \end{center}
\caption{Plot of the wealth distribution for $0 < \gamma \leq 1$ at different times for an
initial normal distribution.} \label{fig:Nwlinmaior} \end{figure}

Condensation of wealth is an effect undesirable in models describing wealth distribution as
well as in real societies. So, many authors proposed different mechanisms in order to avoid
or compensate this effect. One of the most ``popular'' recipes is to define a
wealth-dependent probability of being the winner in the transaction. For example in
refs.~\cite{west,IGVA2004,Caon2007,sinha1,cristian07} it is assumed that the probability of
being the winner in a transaction is higher for the agent with the lower wealth in each
interaction. Even if this recipe seems counterintuitive, increasing the probability of
favoring the poorer agent is a way to simulate the action of the state or of some another
type of regulatory tool aiming to redistribute the resources. In order to verify the effect
of this kind of measure, here we will consider an asymmetric winning probability, by taking
$0 \leq \gamma \leq 1$ in Eq.(\ref{eq:gain}).
The iteration results for a uniform initial distribution are shown in
Fig.~\ref{fig:Nwlinmaior}, where we extended the iteration time from 800 to 3200 timesteps.
The results for other initial conditions are similar, as it was for the $\gamma=0$ case.
However, even considering values of $\gamma \neq 0$, as time evolves the distribution
$N(w)$ tends to an L-shaped distribution where all agents concentrate at $w=0$, with an
infinitesimal number of agents presenting $w>0$: all distributions exhibit peaks for $w=0$
as time increases. The exception happens for $\gamma=1$, when the peak in the distribution
located at $w>0$, is stable and increases as time goes by. In this case the wealth
distribution approaches a shape that is well fitted by a Gaussian or lognormal function.
Nevertheless, it is clear from Fig.~\ref{fig:Nwlinmaior} that for bigger enough values of
$\gamma$ the time for arriving to condensation is also bigger, so, for finite periods of
time, increasing $\gamma$ diminish inequality.

\begin{figure} \begin{center}
\includegraphics[width=8cm]{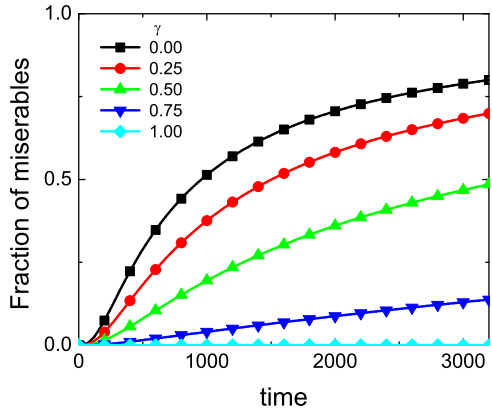} \end{center} \caption{Plot of $N(w=0)$ with time
for normal initial condition and different values of $\gamma$}. \label{fig:n0Xtime}
\end{figure}

The fact that for $\gamma < 1$ the system converges to condensation when $t \rightarrow
\infty$ can be explained because when an agent reaches the miserable state, with $w \cong
0$, it no longer participates in the transactions, since in transactions involving
$w=0$-agents the exchanged amounts are always zero. It means that the $w=0$ state acts as a
trap of zero escaping probability: and it is just a question of time for the system to
reach a state where all agents concentrate at the total misery state. The only case when
this situation does not happen is when $\gamma=1$ because, in this case, the poorer agent
always wins such that the system is driven towards a kind of equipartition of wealth.
Fig.~\ref{fig:n0Xtime} illustrates this point by presenting the evolution of $N(w=0)$ with
time, for different values of $\gamma$. The fraction of miserable agents remains finite, as
in the initial state, only for $\gamma=1$. 

\subsection{Entropy evolution: the second law}
\begin{figure} \begin{center} \includegraphics[scale=0.4,angle=0]{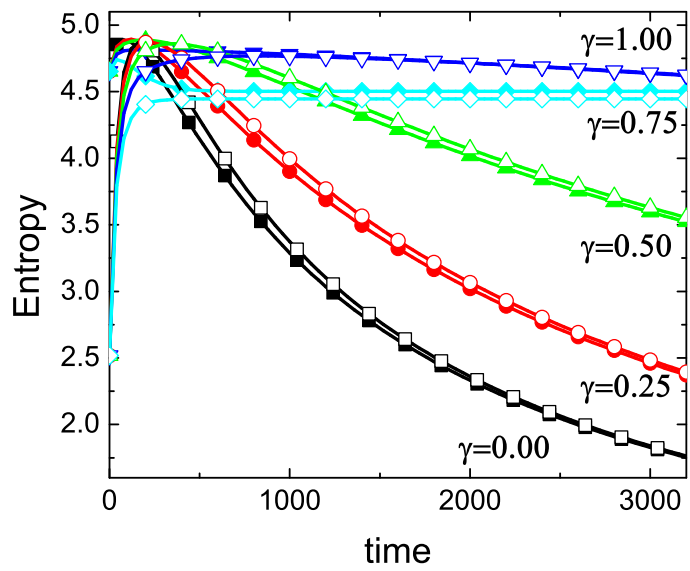}
\end{center} \caption{Time evolution of the entropy for different values of $\gamma$ and
two different initial conditions. } \label{fig:Entropy} \end{figure}

To quantitatively characterize the degree of inequality of the distribution we consider
different estimators. Inequality is generally estimated by either using Gini coefficient~\cite{Gini}, whose calculation is presented at the end of this section, or the
Shannon entropy, known by economists as the Theil coefficient~\cite{Theil}. Here, in order to
compare with a more typical thermodynamical variable, we also calculate the usual entropy
of the system. We begin by calculating the conventional entropy, in the way it is defined
in Statistical Physics textbooks~\cite{Reichl}:
\begin{equation} S=- \int_0^{\infty}N(w,t)\log{N(w,t)}. \end{equation} The time evolution
of the entropy is given by \begin{eqnarray} \frac{\mbox{d}S}{\mbox{d}t}&=&
-\frac{\mbox{d}}{\mbox{d}t} \int_0^{\infty}N(w,t)\log{N(w,t)}\nonumber \\ &=&
\int_0^{\infty}\frac{\partial N(w,t)}{\partial t}\log{N(w,t)} \end{eqnarray} and using
equation \ref{evol1}, we obtain \begin{eqnarray} \frac{dS}{dt}&=& -k \int_0^{\infty}{dw'}
\int_0^{\infty}{dw''}N(w',t)N(w'',t)\nonumber \\ &\times & \log{\left[ \frac{
N\left(w'+\frac{w'w''}{w'+w''},t\right)^{1-\gamma \tanh(w'-w'')}
N\left(w'-\frac{w'w''}{w'+w''},t\right)^{1+\gamma \tanh(w'-w'')}}{N(w't)N(w'',t)}\right] }
\nonumber \end{eqnarray} Even if, for $\gamma=0$, a Boltzmann-like distribution with an
effective temperature proportional to $<w>$, that is, $N(w)
=\frac{\mbox{e}^{-\frac{w}{\langle w \rangle}}}{\langle w \rangle}$, yields
$\frac{\mbox{d}S}{\mbox{d}t}=0$, a visual inspection of Fig.~ \ref{fig:Nwlog} suggests that
this is not the stationary solution for long times even for $\gamma\neq 0$: the wealth
distribution clearly exhibits an exponential behavior for intermediate and low values of
$w$. Alternatively, there exists another type of function that presents norm $1$, average
$\langle w \rangle=1 $, and leads to $\frac{\mbox{d}S}{\mbox{d}t}=0$. This function is also
compatible with the asymptotic stationary solution (as indicated by numerical simulations
for $\gamma<1$), and is a stationary solution of Eq..(\ref{evol1}). One possible function
with these properties type is: \begin{eqnarray} \label{Nstat}
N_{st}(w,t)=\lim_{w_{max}\rightarrow \infty}\left\{ \begin{array}{ll} 1- \frac{2\langle w
\rangle}{w_{max}} & \mbox{ if } w=0 \\ \frac{2\langle w \rangle}{w_{max}^2} & \mbox{ if }
0<w\leq w_{max} \\ 0 & \mbox{ if } w > w_{max} \\ \end{array}\right. \end{eqnarray}
This solution describes condensation that, even if from an economic point of view is the
worst possible scenario, corresponds to the numerical results obtained in this work as well
as in the references quoted above. The solution is also the attractor of the dynamic of the
system: \begin{itemize} \item Due to the dynamics of the system $w_{max}$ is always
increasing. So it tends to infinity for time going to infinity. \item The fraction of the
agents populations with $w=0$ grows as $1- \frac{2\langle w \rangle}{w_{max}} $ and hence
tends to 1 as $w_{max}\rightarrow \infty$. \item The total wealth concentrates in the
infinitesimal population fraction, given by $\frac{2\langle w \rangle}{w_{max}} $ having
$w>0$, which goes to zero as time goes to infinity. \end{itemize} We call this solution
``condensate solution''. The entropy evolution for the iterated distributions is presented
in Fig.~ \ref{fig:Entropy}. For $\gamma<1$ the entropy initially increases, corresponding
to the spreading of the distribution function. However, when the number of agents with zero
wealth increases to very high values, the entropy decreases, indicating an ordered state:
condensation of the agents in the $w=0$ state. This is compatible with the proposed
attractor, Eq.. \ref{Nstat}. It is worth to note that the entropy decreases, and not
increases, in time. We may explain this by the fact that for any initial $w$, an agent has
a non-zero probability of continuously loosing wealth, being attracted to the total misery
condition with $w=0$ when $t \rightarrow \infty$. From that state it is not possible to
escape. Thus, the condensed state plays the role of a zero escaping probability trap in a
random walk and is also an ordered state with minimum entropy.

\section{Minimum entropy: the second law for markets} 
\subsection{Theil entropy} 

The entropy defined in the previous subsection corresponds to the usual Shannon entropy defined
in Physics textbooks when considering the probability of finding an agent with a given
wealth: it corresponds to the integral of the product of the wealth distribution times its
logarithm. An alternative entropy function may be defined considering the probability of
finding a given fraction of wealth with a given agent. To do this we consider the summation
(or the integral) of the wealth times its logarithm, and this is the entropy (or inequality
coefficient) defined by Theil~\cite{Theil} that we will call here Theil entropy, represented
as $S_W$.

\begin{figure} \begin{center} \includegraphics[width=8cm]{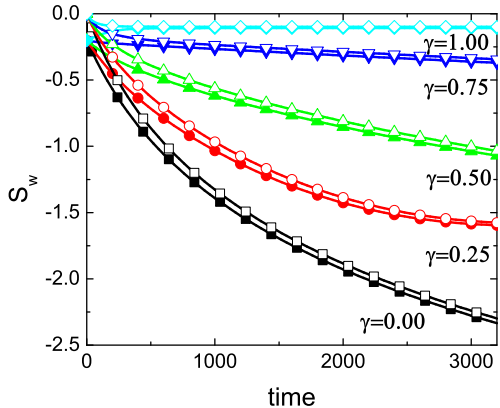} \end{center}
\caption{Time evolution of the entropy for different values of $\gamma$ and two different
initial conditions. } \label{fig:SwXtime} \end{figure}

Consider $W_T$ as the total wealth of the system and $w_i$ as the wealth belonging to agent
$i$. Hence $\frac{w_i}{W_T}$ is the fraction of the total wealth that belongs to agent $i$,
or the probability that a given portion of wealth belongs to agent $i$. This distribution
function is normalized: \begin{equation} \sum_{i} \frac{w_i}{W_T}=1, \end{equation} and we
may define the Theil entropy, $S_W$, regarding this distribution function: \begin{equation}
S_W= - \sum_{i} \frac{w_i}{W_T}\log\left( \frac{w_i}{W_T} \right). \end{equation}
$S_W$ may be calculated using $N(w,t)$ as follows: \begin{equation} S_W=-N_T
\int_0^{\infty} dw N(w,t)\frac{w}{W_T} \log{\left( \frac{w}{W_T} \right)}, \end{equation}
since $W_T = <w> N_T$, where $N_T$ is the total number of agents. Then, the Theil entropy
can be written as: \begin{equation} S_W=-\int_0^{\infty} dw N(w,t)\frac{w}{ \langle w
\rangle} \log{\left( \frac{w}{\langle w \rangle} \right)+ \log(N_T)}, \end{equation} that
is a more appropriate form for numerical purposes.
Fig.~(\ref{fig:SwXtime}) presents the evolution of the Theil entropy for different initial
conditions and values of $\gamma$. Except for $\gamma=1$, $\frac{\mbox{d}S_W}{\mbox{d}t}<0$
for all times. This implies a different second law for the exchanges defined above. This
second law where the entropy decreases instead of increasing characterizes the wealth
concentration process (or condensation) and it is the opposite of the ``equipartition of
energy'' obtained in the kinetic theory of gases. This point will be discussed in detail
below and we will try to built possible physical systems with a similar behavior.

\begin{figure} \begin{center} \includegraphics[width=8cm]{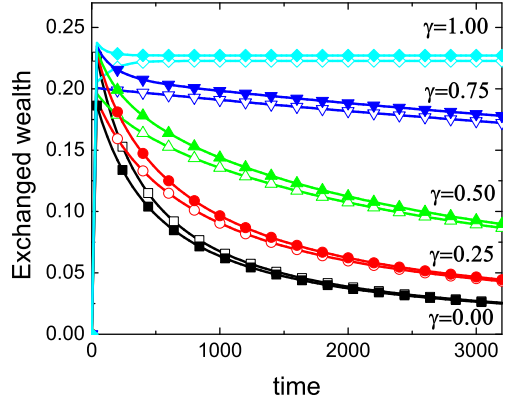} \end{center}
\caption{Evolution of exchanged wealth for different initial conditions and values of
$\gamma$. It is possible to observe that it decreases in time, leading to a thermal death
of the market} \label{fig:ExchXtime} \end{figure}

If the second law of thermodynamics, when applied to the whole universe, has as a corollary
the ``thermal death of the universe'', the concentration - or condensation - of wealth
leads to a ``thermal death of the market'', since the market needs exchanges, or flux of
capital, to survive. If all agents, with a few exceptions, have zero wealth, there is
almost no exchanges. This can be verified if we calculate the ``liquidity'', or the money
being exchanged in the system. We define the liquidity of the market as the amount of money
exchanged per unit time:
\begin{equation} C(t)= \frac{1}{2} \int_0^{\infty} dw'\int_0^{\infty}dw'' N(w',t) N(w'',t)
\frac{w'w''}{w'+w''} \end{equation} The liquidity has been numerically calculated for
different times and it is represented in Fig.~\ref{fig:ExchXtime}. The liquidity is also a
decreasing function of time for $\gamma<1$.

\begin{figure} \begin{center} \includegraphics[width=8cm]{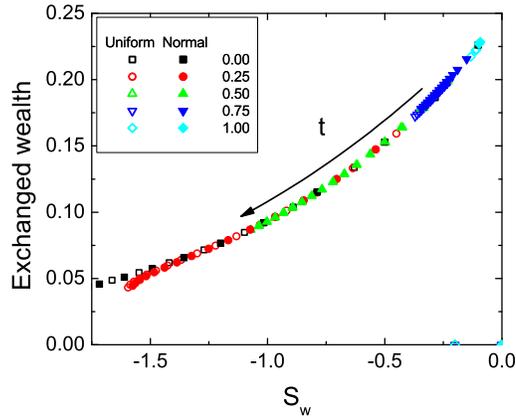} \end{center}
\caption{The liquidity, or circulating capital, $C$, as function of the wealth entropy
$S_W$ for different initial conditions and values of $\gamma$. $C$ increases with $S_W$,
showing that evenly distributed wealth leads to higher wealth exchanges. }
\label{fig:ExcXSw} \end{figure}

One can argue that this behavior of the $W$-entropy and of the liquidity are artifacts
because we have considered a wealth conserving system, and this is not realistic. However,
within a very simplified description of the wealth generation process one can expect that
the wealth production should be proportional to the circulating capital: wealth must be
exchanged to generate more wealth. (Adam Smith said that trade and exchange are emblematic
of the nature of men: ``Nobody ever saw a dog make a fair and deliberate exchange of one
bone for another with another dog.''~\cite{Smith}) In this case it is interesting to
investigate the dependence of the total exchanged wealth, $C(t)$, with the wealth
distribution. Wealth concentration is supposed to decrease with $W$-entropy $S_W$, then, we
represented in Fig.~\ref{fig:ExcXSw} $C$ versus $S_W$ for all times and for both initial
conditions. The circulating capital is an increasing function of $S_W$, which means that
the more concentrated the wealth is, the smaller wealth production should be expected,
confirming that even for a non-conservative market this alternative second law is still
valid.

\subsection{Why the market condenses?}
We focus now on the characteristics of the dynamics that explain condensation as the
attractor of the evolution equation. The first reason is that the state of total misery,
that is, agents with $w=0$, is a trapping state with zero escaping probability. The second
point is that, regardless an agent wealth, the probability that eventually it approaches
$w=0$ is non-null. On the other hand, as wealth is conserved, the accumulation of agents
near $w=0$ implies few agents with very large $w$.
To further illustrate this point, we calculated the average probability $Q_{+}(w)$ for an
agent with wealth $w$ to increase its wealth in a given time step, as \begin{eqnarray}
\label{eq:incprob} Q_{+}(w)&=&\int_{0}^{\infty} dw'\, N(w',t) p_{\gamma}(w|w') \nonumber \\
&=& \frac{1-\gamma <\tanh(w-w')>}{2} \end{eqnarray} where $<...>$ stands for the average
over $N(w',t)$. Assume a wealth distribution peaked around $<w>$. For $w$ much lower than
the typical values in the wealth distribution, $\tanh(w'-w)$ approaches $-1$ when
$N(w',t)>0$ and hence $Q_{+}(w)\sim \frac{1+\gamma}{2}$. If $\gamma<1$ there is a non zero
probability for the agent to loose in a transaction and of approaching the total misery
state. This total misery state is a financial hell, as once there, in the words of Dante:
{\it Lasciate ogni speranza, voi che entrate}, i.e. there are no chances to escape. The
wealth distribution will approach the condensate `L'-shape configuration. Only for
$\gamma=1$ the dynamics guarantees that the probability of loosing wealth is zero for the
poorest. This is the one case where the 'L'-solution is not the attractor of the dynamics.
The stationary regime in this model and the condition of conservation of the total wealth
implies that there are a few infinitely rich agents, since there is not a bound in the
maximum possible wealth. As wealth is conserved that implies many poor agents, and a very
unfair wealth distribution.

\subsection{Lorenz curves and Gini coefficient}
\begin{figure} \begin{center} \includegraphics[width=11cm]{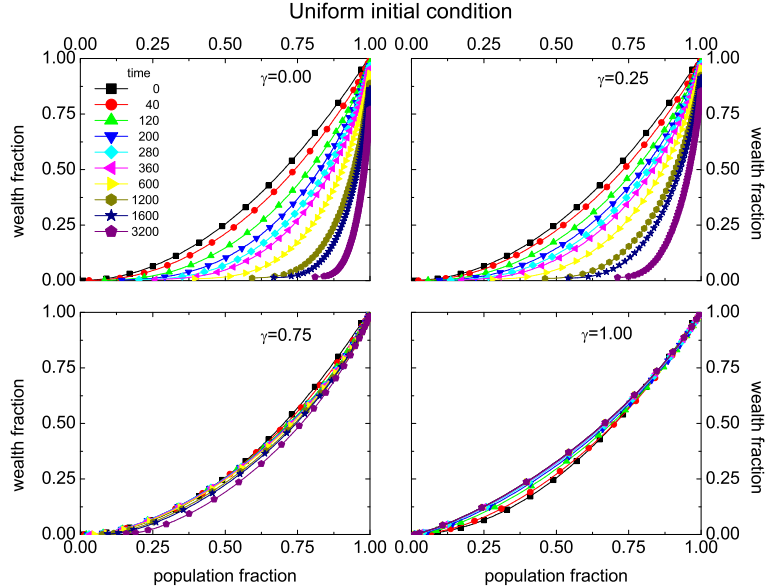} \end{center}
\caption{Lorenz curves as a function of time for different values of $\gamma$, at different
times} \label{fig:Lorenzuni} \end{figure}

In the previous section we have defined the $W$-entropy or Theil coefficient. Another
useful and current measure of inequality is the Gini coefficient~\cite{Gini}. To evaluate
this coefficient, we first construct the Lorenz curves by defining $x(w_{Z})$ as the
fraction of the agent population with wealth lower or equal to $w_Z$, that is,
\begin{equation} x(w_{Z})= \int_0^{w_{Z}} N(w,t) dw, \end{equation} and the fraction of
wealth $F(w_Z)$ belonging to this population as \begin{equation} F(w_{Z})=\int_0^{w_{Z}}
N(w,t)\frac{w}{\langle w \rangle} dw \end{equation} As both $x(w_{Z})$ and $F(w_{Z})$ are
uniquely defined by $w_{Z}$ we may build a function $L(x)$, called the Lorenz curve, as the
fraction of wealth $F(w_{Z})$ calculated at a value of $w_{Z}$ that corresponds to the
population fraction $x$.
Fig.~\ref{fig:Lorenzuni} shows the time evolution of the Lorenz curves corresponding to the
runs with uniform initial conditions. It is clear from the figure that the Lorenz curves
present a negative curvature whose absolute value increases in time for $\gamma < 1$ and,
as one should expect, the limit for $t \rightarrow \infty$ is a condensed state, or an
almost $L$-distribution for zero or a small value of $\gamma$, while the share of the
wealth is rather stable and even improves for high values of $\gamma$.

\begin{figure} \begin{center} \includegraphics[width=8cm]{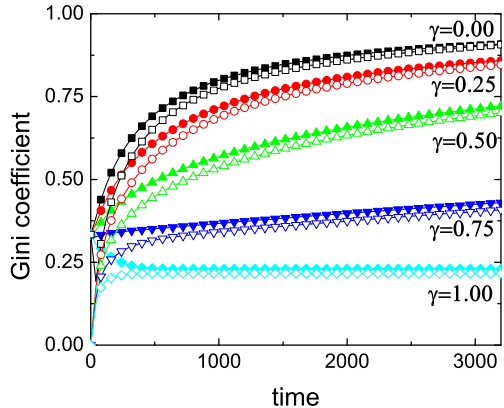} \end{center}
\caption{Time evolution of the Gini coefficient. Note that, after a short transient, the
Gini coefficient is a monotonically increasing function of time when $\gamma < 1$.}
\label{fig:gini} \end{figure}

The Gini coefficient is defined as~\cite{Gini} \begin{equation} G=1-2\int_0^1 L(x)dx
\end{equation} which gives a measure of how unequal the wealth distribution is: for a
distribution where all agents have the same amount of wealth, $N(w,t)=\delta(w-\langle w
\rangle)$, the Lorenz curve is a straight line such that $\int_0^1 L(x)dx=0.5$, yielding
$G=0$. On the other hand when all wealth belong to just one agent, $G=1$. Fig.~
\ref{fig:gini} shows the evolution of the Gini coefficient for the iterations starting from
a normal distribution and from a uniform distribution. The initial value of the Gini
coefficient is higher in the case of the uniform distribution but for $t \simeq 1000$ time
steps the system already ``forgot'' the initial conditions and the Gini coefficient
monotonically increases asymptotically approaching $1$ for $ t \rightarrow \infty$.

\section{Discussion and Conclusion}
We have shown, both by analytical and numerical methods that exchange models where the
exchange fraction cannot be bigger than the capital of any of the participants, leads to a
condensed state and so, to a state of minimum entropy, in disagreement with the second law
of thermodynamics. However, this state is a state of equilibrium and a state that
represents a kind of thermal dead of markets, as no more exchanges are possible when all,
or almost all agents (with the exception of a set of zero measure), have wealth equal to
zero. That means that this kind of exchange rule induces a behavior completely different of the
one predicted by Boltzmann H-theorem. The main difference comes from the {\it quantity} of wealth, (or of the exchanged resource), that the agents can exchange. As we discussed previously, if the
exchange assets are determined completely at random, the Boltzmann-Gibbs distribution is
recovered~\cite{dragu2000}. Also, if an ``unfair'' rule of exchange is introduced, for
instance allowing one agent to take all the assets of the other partner, a distribution
with a higher Gini coefficient is obtained and there is no condensed
state~\cite{hayes,Caon2007}.

Thus, it seems that condensation emerges from the fact that it is impossible to receive more
money that the quantity already owned. What seems to be a fair exchange rule has the
implication of spreading misery. To avoid condensation and the thermal death of markets
some regulation, or some minimum allowance is necessary to favor the poorer agents. When
there are no regulations and/or when no one can win more than he has, the dynamics leads to
a condensed state and to a frozen economy. This result emphasize the importance of
regulations and also of loans, as they permit to invest more capital than the one owned by each
agent. Also, politics of acting on the poorest agent, as described in an extremal dynamics
model~\cite{SI2004}, tend to avoid this condensed state.

If one wants to compare this result with the second law of thermodynamics applied to
physical systems, or to thermodynamic models for economic systems as the ones described,
for example, by Mimkes~\cite{Mimkes} then random or unfair exchanges are essential features
to recover the second law. This effect may seem awkward but we can think of physical
systems where the exchange rule is similar to the one proposed here and, indeed, there
exist physical systems that behave in a similar way. Maybe the best known example are the
{\it moderators} used in nuclear reactors, where particular materials are used to
{\it thermalize} neutrons transforming a wide distribution of energy into a distribution with
a pronounced peak in the region of energy of interest. One can argue that even if the entropy of the neutrons decreases, the total entropy
of the system increases. That is a good clue to better understand the behavior of the
exchange model for markets. If the studied exchange model leads to a decrease of entropy and then
to an ordered state of minimum entropy, this is because a) each agent is described by a single parameter, his wealth, and b) because poor agents have no possibility of recovery. This last item is important when considering policies to improve the wealth distribution. Concerning the first item we should be aware that a pure exchange models provides a limited description of markets and trade, as they do not consider neither regulatory policies, nor production of goods and commodities, nor
salaries, nor banks, nor debts. The limitation of those models was extensively discussed by Gallegati {\it et. al.}~\cite{gallegati}. Nevertheless, in spite of their simplicity, we think that exchange models capture an essential characteristic of economical activity, and, in particular, of markets, i.e. accumulation of wealth in a few hands, on the other hand they show that equilibrium is probably not an essential ingredient in the description of markets. However, in the same way that uniform temperature leads a physical system to thermal death, in economic systems the thermal death
is a consequence of a extreme inequality in revenues and wealth, and we hope to have exhaustively
demonstrated this point in this article.

\section*{Acknowledgements} JRI thanks Viktoriya Semeshenko, Mishael Milakovic, Hugo
Nazareno and Marcel Ausloos for useful discussions and Mar\'{\i}a del Pilar Castillo for a critical reading of the manuscript. The authors acknowledge financial support from Brazilian agencies CNPq and CAPES. \newpage

\end{document}